\documentclass[prl,twocolumn,showpacs]{revtex4}
\usepackage{amsmath}
\usepackage{amsfonts}
\usepackage{amssymb}
\usepackage{graphicx}
\begin{document}
\title{Congestion phenomena on complex networks}

\author{Daniele De Martino$^{a}$ }
\author{Luca Dall'Asta$^{b}$}
\author{Ginestra Bianconi$^{b}$}
\author{Matteo Marsili$^{b}$}

\affiliation{$^{a}$International School for Advanced Studies SISSA and
  INFN, via Beirut 2-4,34014 Trieste, Italy, \\
$^{b}$The Abdus Salam ICTP, Strada Costiera 11, 34014, Trieste, Italy}

\date{\today}

\begin{abstract}
We define a minimal model of traffic flows in complex networks containing the most 
relevant features of real routing schemes, i.e. a trade--off strategy between topological-based
and traffic-based routing.
The resulting collective behavior, obtained analytically for the ensemble of uncorrelated networks, 
is physically very rich and reproduces results
recently observed in traffic simulations on scale-free networks. 
We find that traffic control is useless in homogeneous graphs but may improves global performance in inhomogeneous networks, enlarging the free-flow region in parameter space. Traffic control also introduces non-linear effects and, beyond a critical strength, may trigger the appearance
of a congested phase in a discontinuous manner.
\end{abstract}

\maketitle

The first identified Internet's congestion collapse dates back to October $1986$,
when data troughput from Lawrence Berkeley National Laboratory to the University of California in Berkeley dropped from $32$ Kbps to $40$ bps.
After that initial event, traffic congestion continued to threaten Internet's practitioners, even after the implementation of {\em congestion control} algorithms able to recover the system in case of traffic overloads \cite{J88}. 
Computer scientists have also elaborated several schemes of {\em congestion avoidance}, that should prevent congestion to occur by keeping the system far from high levels of traffic \cite{FJ93}. 
Congestion avoidance and control are performed by continuously updating the dynamics of end-to-end flows in 
response to the variation of the load level in the network. Their functioning depend on the average round-trip-time (RTT) of the Acknowledgement signals (ACKs) used to exchange information between routers. For this reason, the observation of heterogeneous patterns in RTT time series has been often 
proposed as evidence of periods of congestion \cite{C94}. Apart from
these indirect measures, congestion events are difficult to monitor
and study, so that a clear phenomenological picture is still missing. In spite of the wide interest in developing optimal routing algorithms, much less attention is devoted to explore theoretically the dynamical mechanisms responsible of congestion. 
It is possible that a better comprehension of these mechanisms could help in understanding experimental data, in predicting congestion events and designing better routing protocols.

The topological and dynamical properties of distributed information systems, such as the Internet \cite{PV04}, pose theoretical challenges of a similar nature of those addressed in statistical physics. Therefore, understanding network congestion phenomena has become a subject of intense research in this field  \cite{OS98}, in particular after the works by Takayasu and collaborators \cite{TT}, in which the evidence of a phase transition from a free-flow regime to a congested phase depending on the load level was reported.
In a recent work, Echenique et al.  \cite{EGM05} have shown using numerical simulations that the nature of the congestion transition depends on the type of routing rules. They have adopted a routing scheme that could be considered a first approximation of realistic transmission control protocol (TCP) routing: packets follow the shortest path between their source and destination, but small detours are admitted in order to avoid congested nodes.
In case of purely topological routing (e.g. along the shortest paths) they found that the congested phase appears continuously, whereas the transition is discontinuous if some traffic-aware scheme is considered.  
The effect of routing rules on network performance has also been addressed in \cite{TTR04}.

In this Letter, we put forward a minimal model of traffic that preserves all interesting features previously observed in simulations but is simple enough to be studied analytically. Both continuous and discontinuous phase transitions observed in \cite{EGM05} are reproduced, and their relation to microscopic packets dynamics is clarified. 


Let us consider a network of $N$ nodes and let $v(i)$ denote the set of neighbors of node $i$. We describe traffic dynamics as a continuous time stochastic process, in which packets are generated at each node $i$ 
with a rate $p_i$. Each node is endowed with a first-in first-out (FIFO) queue in which packets are stored waiting to be processed. Let $n_i$ be the number of packets in the queue of node $i$. If $n_i>0$, node $i$ attempts to transmit  packets at a rate $r_i$, which represents bandwidth, to one of the neighbors $j\in v(i)$. We assume the following probabilistic routing protocol. First, the node $j$ is chosen at random among the neighbors $v(i)$ of $i$. Second, the fate of the packet being transmitted depends {\em i)} on whether $j$ is the destination node for that packet and {\em ii)} on the state of congestion of node $j$. We model both as probabilistic events: {\em i)} we call $\mu_{j}$ the probability that node $j$ is the destination of the packet being sent, meaning that with probability $\mu_{j}$ the packet is ``absorbed'' in the transfer; and {\em ii)} we assume that the transfer is refused by node $j$ with a probability $\eta(n_j)$, which is increasing with $n_j$; in this case the packet does not leave node $i$. This model relates in a stylized manner {\em 1)} the structural features, encoded in the network and in nodes' characteristics ($p_i$, $\mu_i$ and $r_i$), {\em 2)} the protocol response to traffic, detailed in the function $\eta(n)$ and {\em 3)} the ensuing traffic process.

Our model is a simplifyied version of more elaborate models\cite{GDVCA02,EGM05}. In these, packets are created at each node with a given  
destination node, randomly chosen. Then they are dispatched by a given routing protocol, minimizing some cost along the path. It is reasonable to assume that the statistical nature of the collective behavior does not depend crucially on the details of the routing protocol, and that it can be captured by the much simpler  random diffusion-annihilation process assumed by our mode. In particular, the probabilistic traffic-aware routing assumed above is a simple analytically tractable way to introduce a traffic dependence in packets' dynamics. Also, in shortest-path routing a node of degree $k_i=|v(i)|$ is visited with a probability $\propto k_{i}^{\beta}$, with $\beta \approx 2$, whereas in the random walk protocol assumed by our model the probability of visiting node $i$ is proportional to the degree $k_i$. In both, high degree nodes are more exposed to events of congestion, which is why star-like topological structures are particularly vulnerable to congestion and perform optimally only at low traffic \cite{GDVCA02}. Our model, can easily accommodate for this statistical features with a specific degree dependence of $r_i$ or by considering degree-biased random walks, like in \cite{FF07}. 

The phase transition from free-flow to a jammed phase, is characterized \cite{EGM05} by the order parameter
\begin{equation} 
\rho = \lim_{t \to \infty} \frac{\mathcal{N}(t+\tau)-\mathcal{N}(t)}{\tau P}
\end{equation}
i.e. the percentage of not adsorbed packets for unit time, 
where $\mathcal{N}(t)=\sum_{i}n_i(t)$ is the total number of packets in the system at time $t$, $P=\sum_i p_i$ is the rate of creation of packets and $\tau$ is the observation time. Note that a local order parameter, replacing $\mathcal{N}(t)$ by $n_i(t)$ and $P$ by $p_i$, can be defined in the same way. 

The model discussed above can efficiently be analyzed, in the
asymptotic time limit, within a mean field approximation
$\mathcal{P}(n_{1}, \dots, n_{N}) = \prod_{i}
\mathcal{P}_{i}(n_{i})$. The corresponding master equations can be
solved by message passing-type algorithm \cite{inprep}.



Rather than investigating specific examples, we prefer here to focus on the mechanisms inducing the change from a continuous to a discontinuous phase transition when traffic-aware routing is considered. 
A major insight is obtained rephrasing the problem in terms of ensembles of graphs. We consider uncorrelated random graphs with degree distribution $P(k)$, so that $n_{k}$ represents now the average queue length of nodes in classes of degree $k$. We focus on the simple case $\mu_i=\mu$, $p_i=p$ and $r_i=1$ for all $i$, and the routing protocol $\eta(n)=0$ for $n<n^*$ and $\eta(n)=\bar\eta$ for $n\ge n^*$. The mean-field transition rates for nodes with degree $k$ are $w_{k}(n \to n+1) = p + (1-\mu)(1-\bar{q})\frac{k}{z}[1- \bar{\eta}\theta(n-n^{*})]$ and $w_{k}(n \to n-1) = \theta(n)(1- \bar{\chi})$. Here $z$ is the average degree,  $q_{k}$ is the probability that a node of degree $k$ has empty queue and $\chi_{k}=\bar\eta P\{n_i\ge n^*|k_i=k\}$ is the probability that a node of degree $k$ refuses packets. Likewise $\bar{q} = \sum_{k} q_{k} P(k)$ and $\bar{\chi} = \sum_{k} \frac{k}{z} \chi_{k} P(k)$. \\
The average queue length $n_{k}$ follows the rate equation 
\begin{equation}
\dot{n}_{k} = p +(1-\mu) (1-\bar{q}) \frac{k}{z} (1 - \chi_{k})-(1-q_{k}) (1-\bar{\chi})
\end{equation}
 Note that summing over $k$ and dividing by $p$ we obtain a measure of the order parameter $\rho (p)$. Since $\dot{n}_{k}$ depends linearly on $k$, high degree nodes are more likely to be congested. Therefore, in the stationary state for a given $p$, there exists a threshold real value $k^{*}$ such that all nodes with $k > k^{*}$ are congested whereas nodes with degree less than $k^*$ are not congested. 
Congested nodes ($k > k^{*}$) have $q_{k} = 0$ and $\chi_{k} =\bar\eta$.
For $k<k^*$, the generating function $G_{k}(s) = \sum_{n} P(n_{k})s^n$ of the packets distribution can be computed from detailed balance. This takes the form $G_{k}(s) = q_{k}\left\{ \frac{1-{(a_{k} s)}^{n^{*}}}{1-a_{k} s} + \frac{{(a_{k} s)}^{n^{*}}}{1-(a_{k} - b_{k})s}\right\}$ corresponding to a double exponential, where $a_{k} = [p+(1-\mu) \frac{k}{z} (1-q)]/[1-\bar{\chi}]$ and $b_{k} = \bar\eta [(1-\mu)\frac{k}{z}(1-q)]/[1-\bar{\chi}]$. From the normalization $G_{k}(1)=1$ and the condition $\dot{n}_{k} = 0$, we get expressions for $q_k$, $\chi_k$ and, finally, for $\bar{q}$, $\bar{\chi}$.

The value $k^{*}$ is  self-consistently determined imposing that nodes with $k=k^*$ have $q_{k^{*}} = 0$, $\chi_{k^{*}} =\bar\eta$ and $\dot{n}_{k^{*}}=0$ that translates into the equation
\begin{equation} \label{kstar}
k^{*} = [1-p-\bar{\chi}]/[(1-\mu)(1-\bar\eta)(1-\bar{q})] \\
\end{equation}

\begin{figure}
\includegraphics*[width=0.35\textwidth]{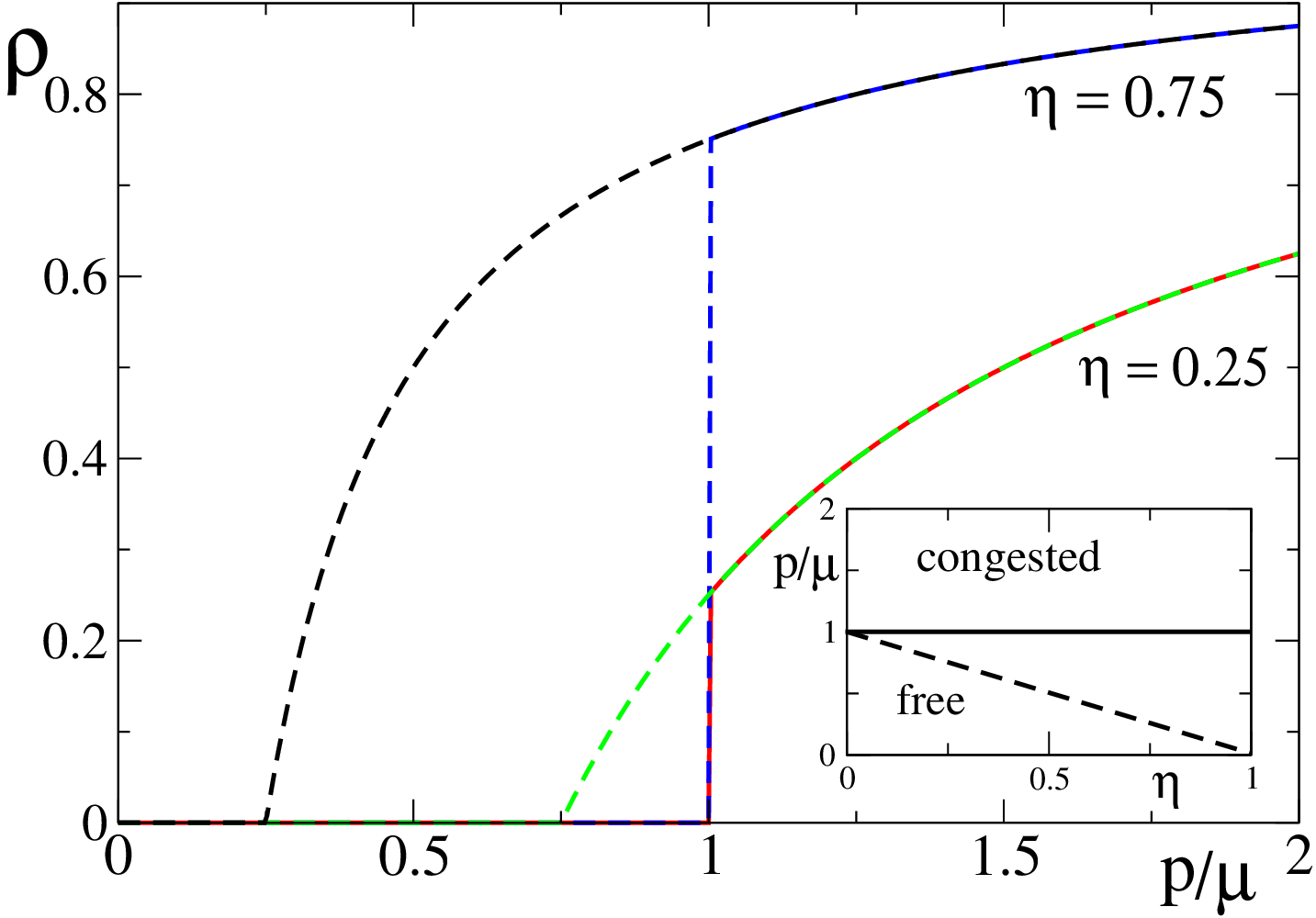}
\caption{
$\rho(p/\mu)$ for an homogeneous graph from theoretical predictions
for $\eta = 0.25$,  $0.75$. Inset:
  phase diagram for the same graph.
}
\label{figreg}
\end{figure}

\begin{figure}
\includegraphics*[width=0.35\textwidth]{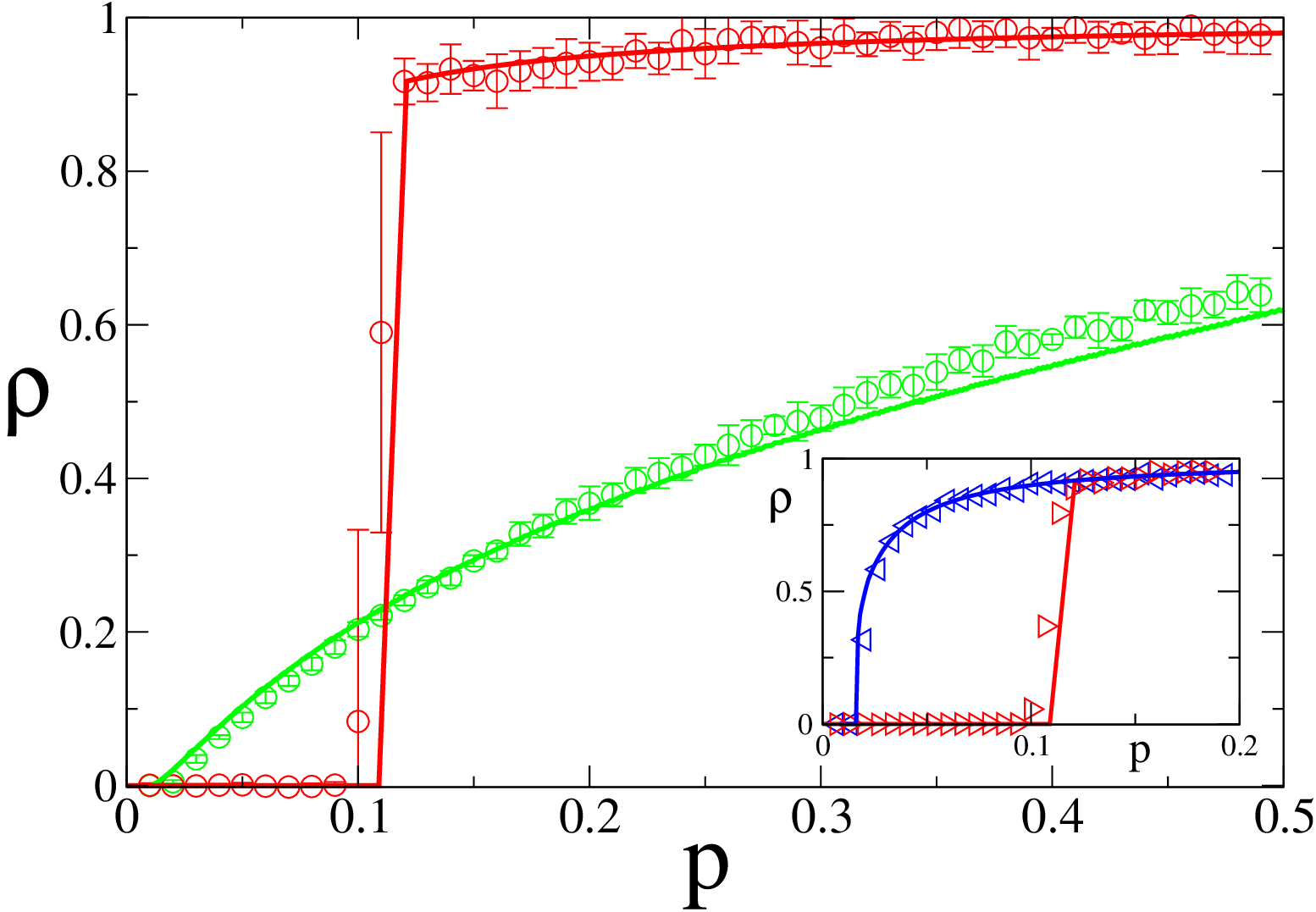}
\caption{
$\rho(p)$ for an uncorrelated scale-free graph ($P(k)\propto k^{-3}$, $k_{min}=3$, $k_{max} = 110$, $N = 3000$), $\mu=0.2$, $n^{*}=10$ and
$\bar\eta=0.05$ (below, green) and $\bar\eta=0.95$ (above, red), from both simulations and theoretical predictions. Inset:
Hysteresis cicle for the same graph for $\bar\eta = 0.95$.
}
\label{fig1}
\end{figure}

The above set of closed equations can be solved numerically for any degree distribution $P(k)$ and $\rho(p)$ can be accordingly computed. 
The solution is particularly simple for regular graphs: $k_i=z, ~\forall i$ in the limit $n^*\gg 1$. Then the congestion free solution with $\rho=0$ has $q_k=\bar q=1-p/\mu$ and $\chi_k=\bar\chi=0$ and it exists for $p\le \mu$. In the congested phase, instead, all nodes have $n_i\to\infty$, i.e. $\bar\chi=\bar \eta$ and $\bar q=0$. This solution has $\rho=\dot n/p=1-(1-\bar\eta)\mu/p$ and exists for $p\ge (1-\bar\eta)\mu$. Therefore, in the interval $p\in [(1-\bar\eta)\mu,\mu]$ both a congested and a free phase coexist, as shown in the inset of Fig. \ref{figreg}. The behavior of $\rho$ as a function of $p$ exhibits hysteresis: The system turns from a free to a congested phase discontinuously as $p$ increases at $p=\mu$ and it reverts back to the free phase from a congested phase at $p=(1-\bar\eta)\mu$ as $p$ decreases. This simple case also shows that traffic control is useless in homogeneous graphs, as it does not enlarge the stability region of the free phase, while making a congested phase stable for $p\in [(1-\bar\eta)\mu,\mu]$ (see inset of Fig. \ref{figreg}). 

\begin{figure}
\includegraphics*[width=0.35\textwidth]{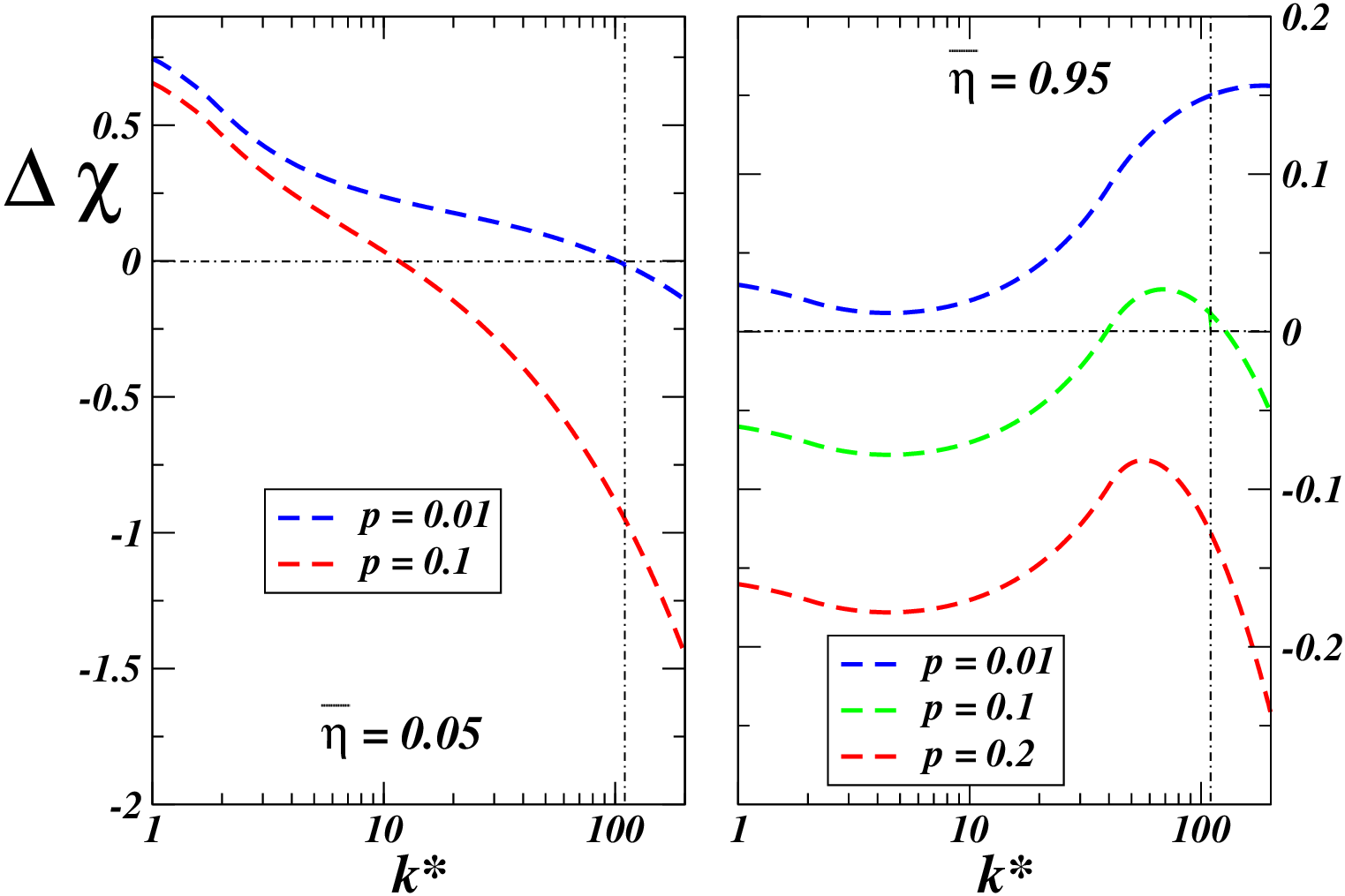}[!h]
\caption{
The zeros of $\Delta \chi (p)$ vs. $k^{*}$ define the threshold degree for the onset of congestion in a network.
The picture refers to a scale-free random network with $\gamma = 3.0$ ($k_{max}= 110$), and different values $\bar\eta = 0.05 ,0.95$ and $p$. The 
solution $k^*_1(p)$ in the right panel falls outside the plot.
} \label{fig2}
\end{figure}

The case of heterogeneous graphs instead is much richer. In Fig. \ref{fig1}, we display $\rho(p)$ for a scale-free network from both simulations (points) and numerical calculations (full line).
The agreement is very good and the behavior of the curves reproduces the scenario observed in \cite{EGM05}. The figure is obtained for $\mu = 0.2$ and $n^{*} = 10$, but the behavior does not qualitatively change for different values of these parameters. The dependence on $\bar\eta$ brings instead qualitative changes. 
Increasing $\bar\eta$ from $0.05$ to $0.95$, the transition becomes discontinuous and $p_{c}$ increases. Fig.\ref{fig1} (inset) also shows that in case of discontinuous transition, the system exhibits hysteresis phenomena: a congested system does not immediately decongest if the creation rate $p$ is decreased under the threshold value $p_{c}$. \\
When the whole system is congested, simple arguments of queueing theory show that $\rho(p)$ follows the curve $1- \frac{(1-\bar\eta)\mu}{p}$; however the most interesting situation occurs when the network is only partially congested.    
This case can be better understood considering the limit $n^{*} \rightarrow \infty$,  in which the calculations simplify considerably without modifying the overall qualitative behavior. \\
In this limit, uncongested nodes have $a_{k} <1$, hence  $\chi_{k} \to 0$ and $q_{k} = 1- a_{k}$, meaning that these nodes have short queues and do not reject arriving packets.   If $a_{k}>1$, then $q_{k} \to 0$ and $\chi_{k} = \bar\eta \frac{a_{k} -1}{b_{k}}$; more precisely we have $\chi_{k} = 1-\frac{k^{*}}{k}(1-\bar\eta)$ for $k < k^{*}$ and $\chi_{k} = \bar\eta$ for $k \geq k^{*}$. The latter class identifies  {\em congested} nodes, while we call {\em fickle} those with $k^{*}(1-\bar\eta) \leq k < k^{*} $. The {\em uncongested} nodes exist up to $k_{F} = k^{*} (1-\bar\eta)$.
Using this classification, we get  a first expression for $\bar\chi$, i.e. $\bar{\chi}_{1}$ $ = 
\sum_{k = k_{F}}^{k^{*}} \left[ 1 - \frac{k^{*}(1-\bar\eta)}{k}\right]\frac{k}{z} P(k) $ $+ \bar\eta \sum_{k=k^{*}}^{k_{max}} \frac{k}{z} P(k)$. 
Eq. (\ref{kstar}) provides a further relation between $\bar{q}$,
$\bar{\chi}$ and $k^*$. $\bar{q}$ can be eliminated using its
definition wich leaves us with an implicit equation for $\bar{\chi}$,
whose solution we call $\bar{\chi_{2}}$. \\
In Fig. \ref{fig2} we plot the difference $\Delta \chi = \bar\chi_{1} - \bar\chi_{2}$ vs. $k^{*}$, for $\bar\eta=0.05$ (left) and $0.95$ (right) and different values of $p$. 
The zeros $\Delta\chi(k^*)$ correspond to the only possible values assumed by $k^{*}$. In both cases, $\Delta\chi$ decreases as $p$. For small rejection probability ($\bar\eta = 0.05$ in Fig. \ref{fig2}), there is only one solution $k^{*}(p)$, which decreases from  $+\infty$ when increasing $p$ from $0$. The value $p_{c}$ at which $k^{*}(p_{c}) = k_{max}$ is the critical creation rate at which highest degree nodes become congested. At larger $p$, $k^{*}(p)$  decreases monotonously until all nodes are congested when $k^{*}(p) = k_{min}$.  Hence for low values of $\bar\eta$, the transition from free-flow to congested phase occurs continuously at the value of $p$ for which $k^{*}(p) = k_{max}$.  \\
At large $\bar\eta$ ($\bar\eta = 0.95$ in Fig. \ref{fig2}), the scenario is much more complex. Depending on $p$, the equation can have up to three solutions, $k^{*}_{1} (p)\leq k^{*}_{2} (p)\leq k^{*}_{3}(p)$. It is easy to check that only $k^{*}_{1}$ and $k^{*}_{3}$ are stable solutions. For $p\ll 1$ there is only one solution at $k^{*}_{3} (p)\gg k_{max}$, corresponding to the free phase. As $p$ increases, two cases are possible: {\em i)} $k^{*}_{3}(p)$ reaches $k_{max}$ before  the local minimum crosses zero, in which case the congested phase emerges continuously; {\em ii)} if instead the solution $k^{*}_{1}(p)$ appears when still $k^{*}_{3}(p) > k_{max}$,  a congested phase appears abruptly.
For sufficiently large values of $p$, only $k^{*}_{1}(p)$ survives, and the network becomes fully congested when $k^{*}_{1}(p) \leq k_{min}$.
Therefore, the existence of a purely discontinuous transition depends strongly on the tail of the degree distribution. \\
In the latter case {\em ii)} the hysteresis phenomenon becomes evident upon varying $p$ in opposite directions: Starting from  the free phase at low $p$, the system selects the solution $k^{*}_{3}(p)$ and follows it upon increasing $p$ until the solution $k_3^*(p)$ disappears. On the contrary, starting from the congested phase (large $p$) the system locks into the congested phase $k_1^*(p)$ and remains congested until the solution $k_1^*(p)$ disappears, with a discontinuous transition (see inset of Fig. \ref{fig1}). Hysteresis also occurs in case {\em i)} for the same reasons. A detailed account of this rich phenomenology will be reported elsewhere \cite{inprep}.  

\begin{figure}
\includegraphics*[width=0.35\textwidth]{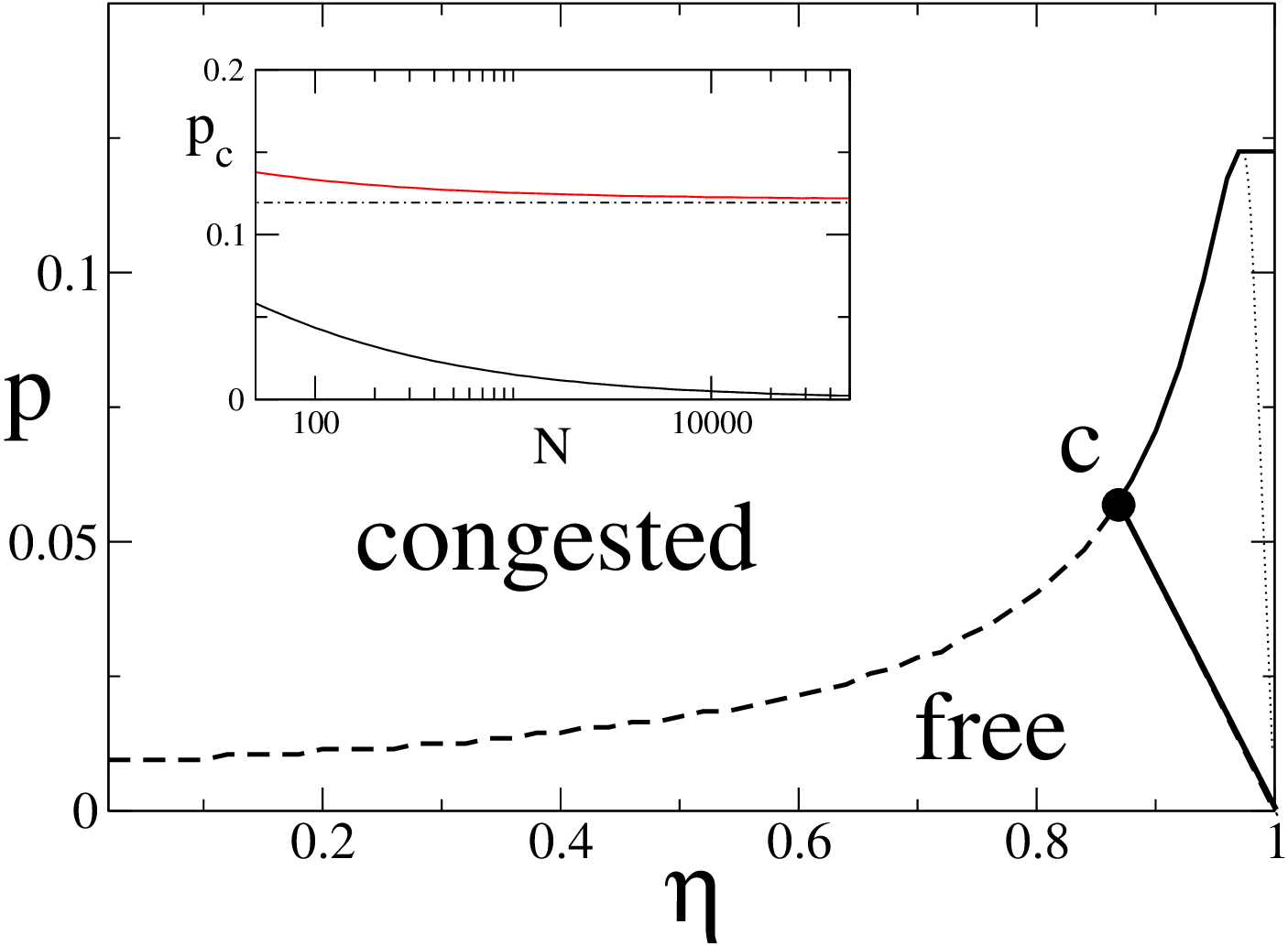}
\caption{
$(\eta,p)$ phase diagram for an uncorrelated scale-free graph ($P(k)\propto k^{-3}$, $k_{min}=3$, $k_{max} = 110$, $N = 3000$), $\mu=0.2$, $n^{*}=\infty$ 
from theoretical predictions. The inset shows $p_{c}(N)$ for $\eta=1$ (red line) and $\eta=0$ (black line).  
} \label{fig3}
\end{figure}

We computed the phase diagram (see Fig. \ref{fig3}) in the plane $(p,\bar\eta)$ for the same uncorrelated scale-free random networks with $\gamma=3$ considered in Fig. \ref{fig1}, in the case $n^* \to \infty$.  The dashed line represents the continuous phase transition, separating free-flow regime from congestion. At the point $C$, the critical line splits in two branches that define a coexistence region. 
The upper full line represents the discontinuous transition from the free-phase to the jammed state, whereas the lower indicates the opposite transition from congestion back to free-flow. The dotted line decreasing from the maximum of the critical line, is an unphysical branch of the solution obtained from calculations. Indeed if a free flow phase is stable for a given value of $\bar\eta$, this will persist for larger values of $\bar\eta$ because traffic control only affects congested nodes in the limit $n^*\to\infty$ (i.e. a free stationary state cannot become congested if we increase $\bar\eta$). 

This phenomenology crucially depends on the tail of the degree distribution. Since $k_{max}$ depends on the system's size, we expect that $p_{c}$ depends on $N$ as well; the inset of Fig.\ref{fig3} show that for the same scale-free network of Fig. \ref{fig2} the critical rate of packets' creation
goes to zero as $p_{c}(N)\propto 1/\sqrt{N}$, for $\eta=0$, but it goes to a constant for $\eta=1$, in the limit of large $N$.  Hence the point $C$ separates two regions in $\bar\eta$ with distinct behavior of finite size effects.

The mechanisms triggering the emergence of congestion is somewhat reminiscent of  jamming or bootstrap percolation,
where a node is occupied if the number of occupied neighbors exceeds a given threshold. In these models, as the threshold increases, the transition turns from continuous to discontinuous \cite{DGM06}. 
Similarly, as traffic control in the routing protocol is enhanced, the congestion transition may turn from continuous to discontinuous, as a
result of a cooperative process in which the more congested neighbors a node has, the more it is likely that it will be congested.\\
In conclusion, we have proposed a minimal model to study the emergence of congestion in information networks.  
For un-correlated random graphs, the analysis can be performed
analytically at the ensemble level, revealing that the interplay between the feedback process induced by traffic-aware routing and the topological structure of the network (in the tail of the degree distribution) generates a rich phenomenology of phase transitions \cite{inprep}. Traffic-aware routing is useful only in heterogeneous networks, where it expands the region of stability of the congestion free state. However, when its effects are strong enough a congested phase may arise abruptly, and once it arises, it may persists even under lower traffic loads.



\end{document}